\documentclass[11pt,aps,floats,showpacs,amssymb,tightenlines,nofootinbib]{revtex4}
\usepackage{color}
\usepackage{graphicx}
\usepackage{amsmath}
\usepackage{amsfonts}
\usepackage{amscd}
\usepackage{epsfig}
\usepackage{amssymb}
\usepackage{tabularx}
\begin{document}

\title[]{Self-gravitating splitting thin shells}
\author{Marcos A. Ramirez}%
\affiliation{Instituto de F\'{i}sica Enrique Gaviola, FaMAF, Universidad Nacional de C\'ordoba, (5000) C\'ordoba, Argentina }%
\begin{abstract}
In this paper we show that thin shells in spherically symmetric spacetimes, whose matter content is described by a pair of non-interacting spherically symmetric matter fields, generically exhibit instability against an infinitesimal separation of its constituent fields.
We give explicit examples and construct solutions that represent a shell that splits into two shells. Then we extend those results for 5-dimensional Schwarzschild-AdS bulk spacetimes, which is a typical scenario for brane-world models, and show that the same kind of stability analysis and splitting solution can be constructed. We find that a widely proposed family of brane-world models are extremely unstable in this sense. Finally, we discuss possible interpretations of these features and their relation to the initial value problem for concentrated sources.

\end{abstract}

\maketitle
\section{Introduction}



Singular surfaces, in the sense of \cite{Israel}, can be of two types: $(i)$ {\it boundary surfaces} or {\it shock waves}, which entail a discontinuity for the stress-energy tensor; and $(ii)$ {\it surface layers} or {\it thin shells}, which entail a distributional stress-energy tensor that represents a concentrated source for the field equations. They are useful to construct models for a number of concrete astrophysical and cosmological situations where shock surfaces form, or where, as a valid simplifying assumption, some part of the system can be considered of vanishing thickness \cite{astrophysics}. 
As in the case of vacuum bubbles and domain walls in the early universe (\cite{phasetransition}, \cite{vilenkin}), singular surfaces are frequently needed to represent interfaces between different matter models, or different {\it phases} of matter. On the other hand, thin shells proved to be a very useful tool for constructing toy-models to gain insight into general theoretical problems in general relativity, like the cosmic censorship conjecture \cite{censor}.

Moreover, there is a relevant family of models in which a zero-thickness configuration is not necessarily a simplifying assumption, and may be interpreted as a fundamental hypothesis. Inspired by string theory, certain cosmological models that consider the visible universe as a part of a singular 4-dimensional surface (thin shell) embedded in a higher-dimensional space-time appeared (\cite{RS1}, \cite{RS2}). They are phenomenological models that can account for the ``invisibility'' of extra dimensions, and constitute a framework where the holographic principle can be naturally understood \cite{Review}. Besides being a theoretical possibility that is worth exploring, brane-world models have been proven to have interesting properties, such as Friedmann-like evolution equations that exhibit an inflationary stage.

In this way, we consider that an analysis of general properties of thin shell solutions is important and may have relevant astrophysical and cosmological implications. As mentioned, for most astrophysical applications, a thin shell constitutes an idealization of an ultimately thick configuration. Therefore, it is relevant to ask whether there actually are {\it smooth} configurations whose evolution in spacetime is somehow ``close'' to a given thin shell solution, as it would be expected if thin shells were to be considered representative idealizations. Related to this, the following question is also relevant: are thin shells solutions {\it stable}? In other words: would an arbitrary configuration that is initially ``close'' to a given thin shell solution, either smooth or non-smooth, remain ``close'' throughout its evolution?

In a previous work \cite{paper1}, it has been shown that, for certain families of self-gravitating thin shells made of collisionless particles, there are configurations
that are identical to a given thin shell solution up to a point where they start evolving in totally different manner. The mechanism to construct these solutions is to infinitesimally separate non-interacting constituents of the ``original'' thin shell and then solve the evolution equations of the resulting spacetime.
It was then said that these families of thin shells, while being well-defined solutions of Einstein equations of weak regularity, result unstable against an infinitesimal separation of constituents, as follows. Taking into account that the constituent particles are supposed to interact only gravitationally, the separation of shell components could only be avoided if these components interact otherwise, so as to counteract the separation, which is driven by gravity.
 Furthermore, these constructions illustrate the lack of uniqueness in the evolution of unstable (in this sense) thin shell solutions because of the fact that, for any given distribution of particles in the angular momentum space and any given initial radius and velocity of the shell compatible with the constraints, a non-splitting solution always exists. Because of the well-posedness of the Einstein-Vlasov system, the evolution of a smooth configuration initially close in some sense to a thin shell solution is unique and it 
should be well-defined at least in a neighbourhood of an initial Cauchy surface. Indeed, in spherical symmetry, as long as there is vacuum in a neighbourhood of the (regular) symmetry center (as it would be the case for a smooth solution ``close'' to any non-collapsing thin shell solution), global existence can be asserted \cite{DafermosRendall}.
In this way, this instability result suggests the absence of smooth solutions that remain close to these unstable families throughout the evolution. We may then regard them as {\it artificial} as they can not properly represent realistic non-zero-thickness configurations.






This work is an extension of the ideas implemented in \cite{paper1} for collisionless matter in a spherically symmetric and $4$-dimensional spacetime to other more general scenarios. Our purpose is to generalize the analysis of the stability against separation of components in order to develop a new tool in the study of thin shell solutions in general, whenever they are needed. We perform this generalization in two different directions: for higher dimensional scenarios and for arbitrary matter models. The relevance of this analysis can be understood as follows. If a thin shell solution is unstable in this sense, this implies the possibility of a construction of a splitting solution with the same initial data than that of the original thin shell solution. Furthermore, this splitting solution can be interpreted as the result of a ``perturbation'', as it would be the consequence of an infinitesimal displacement of the components at a given time and their further evolution according to Einstein equations. This two facts might compromise the adequacy of the model involving a single and multicomponent shell. On the other hand, it might also imply, as discussed in Section \ref{final}, that this unstable thin shell solution can not be a limiting case of thick configurations.


More precisely, we want to investigate whether thin shells made of a pair of non-interacting constituents entail instability, and hence non-uniqueness in the evolution, by constructing well-defined spacetimes where a single thin shell splits into a number of sub-shells, each one made of one the original constituents, in a differentiable manner.
We particularly focus on brane-cosmology scenarios, where we show that this kind of splitting turns out to be possible, and a wide family of brane-world solutions results unstable in this sense. 

\subsubsection*{Outline}

We begin with a description of a general shell in an isotropic (spherical, planar or hyperbolic symmetry) bulk spacetime of an arbitrary number of dimensions in Section \ref{thinshells}.
Later in Section \ref{2splitting} we develop the stability analysis against separation of non-interacting constituents in spherical symmetry, find a stability condition for general constituents, and then construct an example of a splitting solution.
Then in Section \ref{splittingbranes} we extend the stability analysis to brane-world models. In Section \ref{final} we comment on the relation between the results of this paper and previous works, and possible interpretations of these features in the context of the general initial value problem with concentrated sources. Finally, in Section \ref{concluding} we summarize the content of this work and give some concluding remarks.

\section{Thin shells in $D=n+2$ dimensions}
\label{thinshells}

We begin with a description of a singular shell embedded in an isotropic 
Lambda-vacuum spacetime of an arbitrary number of dimensions. We consider a $n+2$-dimensional spacetime where there is a foliation of $n+1$-dimensional space-like surfaces which posses a kind of $n$-dimensional isotropy: hyperbolic, planar or spherical symmetry.
There is a singular timelike orientable hypersurface $\Sigma$ embedded in the bulk spacetime. Because of the symmetry, the surface must be defined in the space of group orbits. 
In particular, in gaussian coordinates adapted to the symmetry, in a neighborhood of $\Sigma$ the metric reads,
\begin{equation}
ds^2=-f(\tau,\eta)d\tau^2+d\eta^2+a(\tau,\eta)^2\left[\frac{d\chi^2}{1-k\chi^2}+\chi^2d\Omega_{n-1}^2 \right]
\end{equation}
where $\eta=0$ characterizes the surface, $f(\tau,0)=1$ ($\tau$ is the shell proper time), and $k=-1,0,1$ is the curvature index of the group orbits. Also in these coordinates, the intrinsic metric takes the form,
\begin{equation}
\label{intrinsic}
ds^2_{\Sigma} = -d\tau^2 + R(\tau)^2 \left[\frac{d\chi^2}{1-k\chi^2}+\chi^2d\Omega_{n-1}^2 \right]
\end{equation}
where $R(\tau)\equiv a(\tau,0)$. This expression illustrates the fact that any timelike surface defined in the quotient manifold of the group orbits is a FRW submanifold of dimension $n+1$.



In the case of spherical symmetry ($k=1$) and $\Lambda=0$, by virtue of Birkhoff theorem, vacuum solutions of Einstein equations can always be written in the form (see, for example, \cite{Tangherlini}),
\begin{equation}
\label{schwarzschild}
ds^2=-F(r)dt^2+F(r)^{-1}dr^2+r^2d\Omega_n^2
\end{equation}
where $F(r)=1-2M/r^{n-1}$ and $M$ is a mass parameter proportional to the ADM mass.  
On the other hand, Bowcock, Charmousis and Gregory \cite{BCG} have shown 
that in the case $n=3$ this kind of Lambda-vacuum spacetimes
are always 5-dimensional Schwarzschild-Anti de Sitter with the corresponding $k$ parameter. In this way, in the context of SMS brane-world models, we can express the metric in any bulk region with
\begin{equation}
\label{bulkmetric}
ds^2=-F(r)dt^2+F(r)^{-1}dr^2+r^2\left[\frac{d\chi^2}{1-k\chi^2}+\chi^2d\Omega^2 \right]
\end{equation}
where $F(r)=k-2M/r^2+r^2/\ell^2$, and $\ell^2\equiv-6/\Lambda$.


Einstein equations imply junction conditions on the surface \cite{Israel} that relate the jump of the surface extrinsic curvature with the effective mass-energy tensor on the shell. These are the so-called Darmois-Israel junction conditions. With these expressions, for both the spherically symmetric $n+2$-dimensional case and the $5$-dimensional Schwarzschild-Anti de Sitter case, we can write the extrinsic curvature on a given {\it side} of the shell in terms of $R(\tau)$ and the function $F(r)$ that characterizes the bulk spacetime there
\begin{equation}
\label{extrinsic2}
K^i_j = \mbox{sign}\left(\left.\frac{\partial r}{\partial \eta}\right|_{\eta=0}\right)\mbox{diag}
\left[\frac{F'(R)+2\ddot{R}}{2\sqrt{\dot{R}^2+F(R)}},\frac{\sqrt{\dot{R}^2+F(R)}}{R},..,\frac{\sqrt{\dot{R}^2+F(R)}}{R}\right].
\end{equation}
In this way, giving $F(r)$ and specifying whether $r$ increases or decreases with $\eta$, we get an expression for the extrinsic curvature in terms of the {\it intrinsic} function $R(\tau)$.

On the other hand, the matter content of the shell is typically described by a tensor $S^i_j$ defined on $\Sigma$ such that we can formally write the $D$-dimensional stress-energy tensor as,
\begin{equation}
T^a_b = \delta(\Sigma) S^a_b
\end{equation}
where the $S$ tensor in shell coordinates takes the form
\begin{equation}
\label{S}
S^i_j = \mbox{diag} [-\rho(\tau),p(\tau),...,p(\tau)].
\end{equation}
So, as a result of the symmetry imposed, we can describe the matter content of the shell as if it were composed of an $n$-dimensional perfect fluid, whose flow lines follow the trajectories of the comoving observers. Explicitly,
\begin{equation}
S^{ij}=p h^{ij} +(\rho + p)u^iu^j
\end{equation}
where $h_{ij}$ is the intrinsic metric defined in (\ref{intrinsic}), and $u^i=(\partial/\partial\tau)^i$. If there is not hysteresis, we should be able to write $\rho$ and $p$ as functions of $R$. In that case, conservation of the source would read
\begin{equation}
\label{conservation}
\frac{d\rho}{dR}+\frac{n(\rho+p)}{R}=0.
\end{equation}
This equation together with an equation of state $f(\rho,p)=0$, provided it exists, determine $\rho(R)$ and $p(R)$.


Now we can see that Darmois-Israel junction conditions should relate $R(\tau)$, it first two derivatives, and the parameters $M$ (and $k$ and $\Lambda$ in the brane-world case) with the matter functions $\rho(R)$ and $p(R)$. Concentrated matter on the shell implies the {\it discontinuity} of the extrinsic curvature. Looking at (\ref{extrinsic2}), we realize that the jump should be ascribed to a difference between the mass parameters or the cosmological constants for the empty regions at both sides of the shell, which we call ($M_I$,$\Lambda_1$) and ($M_{II}$,$\Lambda_2$), and, eventually, to different signs for $\partial r/\partial \eta$ at both sides. For simplicity, we will set $\Lambda_1=\Lambda_2=\Lambda$\footnote{For an analysis in the case of different cosmological constants at each side of the bulk see \cite{DeruelleDolezel}.}.
In this paper we consider situations where at least one of the two 
regions, say region $I$, is {\it interior}, that is, it can be described by an inequality $r<R(t)$ in terms of the standard coordinates for that region. The reason for this choice, which only precludes the case where both regions are {\it exterior}, will become clear when we analyze the junction conditions.
From now on, we choose the $\eta$ coordinate to decrease when going into region $I$, which implies $\partial r/\partial \eta|_{\eta=0^-}>0$. The junction conditions then read,



\begin{eqnarray}
\label{Israelndima}
\frac{n}{R}\left(\pm\sqrt{\dot{R}^2+F_{II}}-\sqrt{\dot{R}^2+F_{I}}\right) &=&- \kappa \rho \nonumber \\
& & \\
\label{Israelndimb}
\pm\frac{F'_{II}+2\ddot{R}}{2\sqrt{\dot{R}^2+F_{II}}} - \frac{F'_{I}+2\ddot{R}}{2\sqrt{\dot{R}^2+F_{I}}}+\frac{n-1}{R}\left(\pm \sqrt{\dot{R}^2+F_{II}}-\sqrt{\dot{R}^2+F_{I}}\right)&=& \kappa p \nonumber \\
& &
\end{eqnarray}
where 
$\pm=\mbox{sign}(\partial r/\partial \eta|_{\eta=0^+})$.
The situation not considered by these expressions, the case where $I$ and $II$ are both {\it exterior} regions, is not generally regarded as {\it physical} because it would imply a {\it negative energy density} $\rho$.
In the context of brane cosmology, however, a {\it brane tension} is assumed, which implies an effective constant energy density on the shell that could in principle be negative, as it is a non-dynamical entity.
But, if that were the case, we would not recover standard cosmology as a low energy (at least since nucleosynthesis) limit. So even in those speculative scenarios a negative energy density type of matter-energy is hard to justify phenomenologically. In this way, we will only consider matter that satisfies the dominant energy condition, which is a prescription that includes an eventual positive brane tension.
In particular, in the case of a {\it exterior} region $II$, a positive effective energy density $\rho$ would imply $M_{II}>M_{I}$.

From (\ref{Israelndima}) we can obtain an equation of motion that results independent of the sign $\pm$ and reads
\begin{equation}
\label{eqnmotion1}
\frac{1}{2}\dot{R}^2+ V(R) = 0 \; \; \;\; , \;\; \; \; V(R)= \frac {1}{4}(F_I + F_{II}) - \frac{n^2(F_{I}-F_{II})^2}{8\kappa^2 \rho^2 R^2}-\frac{\kappa^2 \rho^2 R^2}{8 n^2}.
\end{equation}

In the brane cosmology setting, this equation can be written as follows,
\begin{equation}
\label{BranesinZ2}
H^2\equiv\left(\frac{\dot{R}}{R}\right)^2 =\frac{\kappa^2 \rho^2}{36}+ \frac{9(M_{II}-M_I)^2}{\kappa^2\rho^2R^8}+\frac{M_I+M_{II}}{R^4}-\frac{1}{\ell^2}-\frac{k}{R^2}.
\end{equation}
In order to recover the classical Friedmann equations as a low energy limit, brane tension is needed ($\rho=\rho_m+b,p=-b+p_m$). In that case, (\ref{BranesinZ2}) can be written \cite{Stoyca},
\begin{equation}
\label{FriedmannsinZ2}
H^2=\frac{\cal{C}}{(\rho_m/b+1)^2R^8}+ \frac{8\pi G}{3}\rho_m\left(1+\frac{\rho_m}{2b} \right) -\frac{k}{R^2} + \frac{M_I+M_{II}}{R^4} +\frac{\Lambda_4}{3}
\end{equation}
where $G\equiv \kappa^2 b/48\pi$, $\Lambda_4 \equiv (\kappa^2 b^2+6\Lambda_5)/12$, and $\cal{C}$ $\equiv 3 (M_{II}-M_I)^2/16\pi G b$.
We notice that there is a ``dark radiation'' term and a term related to the ``asymmetry'' of the bulk spacetime which may cause an inflation-like stage at early ($\rho_m>b$) times\footnote{Which nevertheless cannot account for actual inflation, as explained in \cite{DavisDavis}.}.

On the other hand, from equation (\ref{Israelndimb}) we can obtain for the case of $n+2$-dimensional spherical symmetry,
\begin{equation}
\label{accel1}
\ddot{R}=-\frac{n-1}{2R}\left[1+\dot{R}^2 \mp \left(\frac{2n}{n-1}\alpha+1\right)\sqrt{1+\dot{R}^2-\frac{2M_I}{R^{n-1}}}\sqrt{1+\dot{R}^2-\frac{2M_{II}}{R^{n-1}}} \right]
\end{equation}
where $\alpha\equiv p/\rho$. Analogously, for the brane-world case (\ref{Israelndimb}) would imply
\begin{equation}
\label{accelbrane2a}
\ddot{R}=-\frac{1}{R}\left[k + \frac{2R^2}{\ell^2}+\dot{R}^2 \mp \left(3\alpha+1 \right) \sqrt{\dot{R}^2+F_I}\sqrt{\dot{R}^2+F_{II}} \right].
\end{equation}
These expressions will be useful later on.

\section{Stability of a two-component hypersurface against separation of the constituents}
\label{2splitting}

A general stability analysis should take into account every possible and sufficiently ``small'' variation around an initial data set associated to a given solution, and determine whether the evolution of the ``perturbed'' data remains ``close'' to the aforementioned solution throughout this evolution. Results of this kind are extremely difficult to obtain.
Nevertheless, a partial analysis that takes into account only certain type of perturbations can be useful to gain insight into the general problem. In particular, the appearance of instabilities in any of these partial analysis may imply general instability. 

As mentioned in the introduction, in a previous work \cite{paper1} different kinds of stability against separation of constituents were considered for a spherical self-gravitating thin shell made of Vlasov matter. After an analysis of the stability of test particles moving (initially) within the shell, it was natural to ask what could be the consequences for the stability if we infinitesimally separate the constituents particles into two groups, instead of separating a ``single particle''. 
In particular, for shells whose constituent particles have two possible values for their angular momentum, the instability against separation of the two group of particles naturally defined in this setting was considered.      
It was then obtained that this type of stability analysis is not trivial as it puts constraints on the parameters of the shell. Depending on these parameters, there are solutions that are always unstable, solutions that are always stable, and solutions that are initially stable but become unstable at certain point of their evolution. In this spirit we now consider a gene\-ral\-ization of this analysis to shells composed of general and self-gravitating, but non-interacting, matter fields. Consider for simplicity a shell composed of two non-interacting spherically symmetric matter fields. As we saw above, the symmetry implies that their flow lines are tangent to the comoving world-lines within the shell and that they can be characterized as $n$-dimensional perfect fluids. The infinitesimal separation of the components takes place at a given proper time $\tau_0$, and the relevant information regarding stability is the relative acceleration that the separating shells would have. The resulting spacetime, once both components are infinitesimally separated, can be determined merely by imposing continuity of $R_0\equiv R(\tau_0)$ and its first derivative, which is equivalent to imposing smoothness (each ``branch'' should be an embedded submanifold) in the geometry of each ``splitting'' component, and also equivalent to imposing continuity of normal vectors at the separation point.

The stress-energy tensor takes the form,
\begin{equation}
\label{twocomponent}
S^{ij} = S_1^{ij} + S_2^{ij} = (p_1+p_2) h^{ij} +(\rho_1+ \rho_2 + p_1 + p_2)u^iu^j
\end{equation}
Conservation of the source holds separately for each component,
\begin{equation}
\frac{d\rho_i}{dR}+\frac{n(\rho_i+p_i)}{R}=0
\end{equation}
so we have in principle functions $\rho_i(R)$ and $p_i(R)$ which are characteristic of the matter models.
We solve the equation of motion for the composed stress-energy tensor (\ref{twocomponent}) and then for each component separately, taking into account the geometry that the new configuration would have, as illustrated in Fig.1. {\it If the resulting relative acceleration turns out to have the same sign as the relative displacement, then the shell is unstable, otherwise it should be considered as stable}.

In this Section we prove that certain families of shells, that are solutions of Einstein equations, are unstable against this kind of separation, while other families are stable. The unstable families naturally include the ones previously obtained in \cite{paper1} for Vlasov matter. Nevertheless, the derivation of these instabilities does not decide whether there are solutions that are initially stable and evolve as a single shell up to a point where they \textit{become} unstable and consequently could \textit{undergo} splitting.  
In \cite{paper1}, it was shown by providing explicit examples that such situation is possible for shells made of Vlasov matter. At the end of this Section, we consequently give a further example of a splitting solution for the case of two non-interacting barotropic fluids. 


The keys to analyse the relative acceleration of the resulting single-component shells are equation (\ref{accel1}) and Birkhoff's theorem.
Because of this theorem, the spacetime region that would appear between the splitting shells is characterized by a mass parameter $M_{int}$.
For definiteness, we label with the index 2 ($\rho_2$,$p_2$) the splitting shell that moves {\it into} region $II$, and with the index 1 the other one. 
The equation of motion for the shell labeled by the index $1$ is obtained by replacing $(F_I,F_{II},\rho(R))$ in (\ref{eqnmotion1}) with $(F_I,F_{int},\rho_1(R))$;
and for the shell labeled by the index $2$ by replacing the former quantities with $(F_{int},F_{II},\rho_2(R))$.
In this way, the continuity of $\dot{R}(\tau_0)$ at the separation point determines $M_{int}$ as a function of $(R_0,M_I,M_{II})$ and the matter model parameters.

\begin{figure}
\label{splitting2}
\centerline{\includegraphics[height=8cm,angle=0]{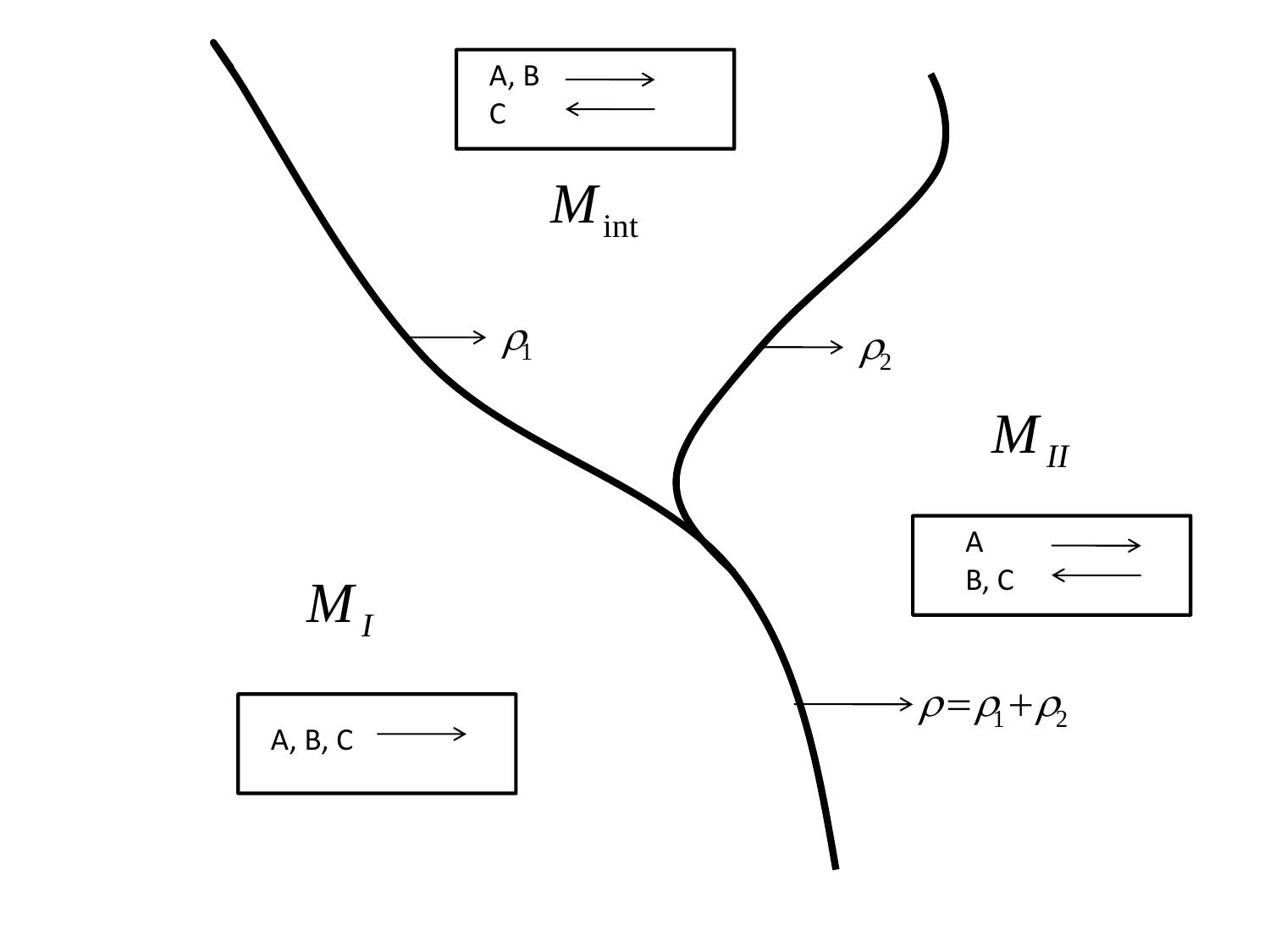}}
\caption{Schematic representation of a {\it splitting} into two parts. The arrows inside the boxes represent the possible orientations of $\nabla^a r$ in each region. We distinguish the three possible sets of orientations (which we call $A$, $B$ and $C$) compatible with the positivity of $\rho_1$ and $\rho_2$.}
\end{figure}

Looking at (\ref{Israelndima}), it can be seen that positivity of $\rho_1$ and $\rho_2$ implies that $M_{int}$ should be greater than at least one of the given mass parameters.
In particular, it implies that if region $II$ is {\it exterior}, then $M_{I}<M_{int}<M_{II}$, and $\nabla_a r$ in the intermediate region should point to region $II$ (we label this possibility for the orientations with the $A$ letter).
In the same way, if region $II$ is {\it interior}, then $M_{int}>M_{II}>M_I$, and both orientations for $\nabla_a r$ in the intermediate region are possible (we call $B$ the case where $\nabla_a r$ points to $II$, and $C$ the case where it points to $I$).


As we said, we can distinguish three possible splitting scenarios, depending on the character of region $II$ and the region that would form between the separating components, as illustrated in Fig.1. We want to compute for each situation the relative acceleration of the shells, and to achieve this we propose using (\ref{accel1}) adapted to each shell as follows,
\begin{eqnarray}
\ddot{R_1}(\tau_0) &=&-\frac{n-1}{2R_0}\left[1+\dot{R}^2(\tau_0)-\mbox{sign}\left(\left.\frac{\partial r}{\partial \eta_1}\right|_{int} \right) \left(\frac{2n}{n-1}\alpha_1+1\right) \right. \mbox{...} \nonumber \\
&& \mbox{...} \left.\sqrt{1+\dot{R}^2(\tau_0)-\frac{2M_I}{R_0^{n-1}}}\sqrt{1+\dot{R}^2(\tau_0)-\frac{2M_{int}}{R_0^{n-1}}} \right] \nonumber \\
\label{accelsplit1}
\ddot{R_2}(\tau_0) &=& -\frac{n-1}{2R_0}\left[1+\dot{R}^2(\tau_0)-\mbox{sign}\left(\left.\frac{\partial r}{\partial \eta_2}\right|_{II} \right) \mbox{sign}\left(\left.\frac{\partial r}{\partial \eta_2}\right|_{int} \right)  \left(\frac{2n}{n-1}\alpha_2+1\right)\right. \mbox{...} \nonumber  \\
&& \mbox{...} \left. \sqrt{1+\dot{R}^2(\tau_0)-\frac{2M_{int}}{R_0^{n-1}}}\sqrt{1+\dot{R}^2(\tau_0)-\frac{2M_{II}}{R_0^{n-1}}} \right]
\end{eqnarray} 
where $\alpha_i\equiv p_i/\rho_i$. We call $\tau$ the proper time coordinate of the original shell and ($\tau_1$, $\tau_2$) those of the resulting shells, and make them coincide at the splitting point $\tau=\tau_1=\tau_2=\tau_0$.

These derivatives are with respect to different time coordinates, so in order to make a comparison we should rewrite them in terms of a single time.
A natural choice in a splitting scenario would be the standard time coordinate of the Schwarzschild intermediate region $t_{int}$. To compute those derivatives we consider the following. For a general Schwarzschild spacetime that has a shell as a boundary, and whose related variables and coefficients we note by a tilde, we can write,

\begin{equation}
\label{taut}
\left(\frac{d\tau}{d\tilde{t}}\right)^2=\frac{\tilde{F}^2}{\tilde{F}+\dot{R}^2}
\end{equation}
where $\tau$ is the proper time coordinate of the boundary surface, $R$ its radius, and $\tilde{F}\equiv 1-2\tilde{M}/R^{n-1}$.
This expression implies that in a splitting scenario we have $d\tau_1/dt_{int}=d\tau_2/dt_{int}$ at $\tau_0$.
In this way, at a given proper time $\tau_0$ we can write,
\begin{equation}
\label{Rtt}
\frac{d^2R(\tau_0)}{d\tilde{t}^2} = \frac{1}{\tilde{F}}\left(\frac{d\tau}{d\tilde{t}}\right)^4\ddot{R}(\tau_0) + \frac{\tilde{F}'}{2\tilde{F}}\left(\tilde{F}-\left(\frac{d\tau}{d\tilde{t}}\right)^2\right)\left(2\tilde{F}-\left(\frac{d\tau}{d\tilde{t}}\right)^2\right)
\end{equation}
which in the splitting scenario implies
\begin{equation}
\left.\left(\frac{d^2R_2}{dt_{int}^2}-\frac{d^2R_1}{dt_{int}^2}\right)\right|_ {\tau=\tau_0}=\frac{F_{int}(R_0)^3}{(F_{int}(R_0)+\dot{R}(\tau_0)^2)^2}\left.\left(\frac{d^2R_2}{d\tau_2^2}-\frac{d^2R_1}{d\tau_1^2}\right)\right|_{\tau=\tau_0}.
\end{equation}

We now can see that, as a consequence of the continuity of $R(\tau)$ and it first derivative, the stability condition turns out to be simply
\begin{equation}
\label{unstable}
\mbox{sign}\left.\left(\frac{\partial r}{\partial \eta_1}\right|_{int} \right)(\ddot{R_2}(\tau_0)-\ddot{R_1}(\tau_0))< 0.
\end{equation}
Taking into account (\ref{accelsplit1}) and (\ref{Israelndima}) for each splitting shell, this inequality takes the form
\begin{equation}
\label{instability1}
\beta \equiv \kappa^2(\rho_1+\rho_2)^2 R^{n+1}(n(\alpha_1+\alpha_2+1)-1)-2n^3(M_{II}-M_I)(\alpha_2-\alpha_1)>0
\end{equation}
which is independent of the signs of $\partial r/\partial \eta|_{int}$ and $\partial r/\partial \eta|_{II}$. It is remarkable that {\it in all the three cases $A$, $B$ and $C$, the stability condition takes the form (\ref{instability1})}. As a first result, we note that this condition implies that \textit{if both fluids are identical ($\alpha_2=\alpha_1=\alpha$) and $\alpha >-(n-1)/2n$, then the shell is always stable}.
A somehow unexpected result is that if we have two identical fluids with $\alpha<-(n-1)/2n$, then the system is unstable. That would mean that if the matter content of a shell satisfies the latter inequality and it makes sense to separate the stress-energy tensor into two non-interacting equal-alpha parts, then such a shell would always be unstable.

Let us assume that both $\alpha_i \geq 0$ (which is the case for matter made of particles), then there is instability only if $\alpha_2$ is large enough in comparison to $\alpha_1$. In this case, if $\rho^2R^{n+1}>(2n^2/\kappa^2)(M_{II}-M_I)$, then (\ref{instability1}) implies that the shell is stable. On the other hand, if $\rho^2R^{n+1}<(2n^2/\kappa^2)(M_{II}-M_I)$, which must hold if $R$ is large enough, (\ref{instability1}) would imply an upper bound for $\alpha_2$ given by
\begin{equation}
\alpha_2 < \frac{(n-1)\kappa^2(\rho_1+\rho_2)^2 R^{n+1}+(n\kappa^2(\rho_1+\rho_2)^2 R^{n+1}+2n^3(M_{II}-M_I))\alpha_1}{2n^3(M_{II}-M_I)-n\kappa^2(\rho_1+\rho_2)^2 R^{n+1}}.
\end{equation}
For large $R$, this upper bound tends to $\alpha_1$. Considering that the analysis is equally valid if we interchange the indexes $1$ and $2$, this implies that if the asymptotic values of $\alpha_1$ and $\alpha_2$ do not coincide (if $\alpha_2-\alpha_1$ grows with $R$ or tends to a non-zero constant) then the shell turns out to be unstable for $R$ large enough. 


\subsubsection*{\textbf{Fluids with equation of state $p_i=\omega_i\rho_i$}}

We now evaluate the stability criterion in the case of 
two fluids with equation of state $p_i=\omega_i\rho_i$, where $\omega_2>\omega_1$. In this case (\ref{instability1}) takes the form
\begin{equation}
(C_1R^{-n(1+\omega_1)}+C_2R^{-n(1+\omega_2)})^2 R^{n+1} (n(\omega_1+\omega_2+1)-1) - 2n^3(M_{II}-M_{I})(\omega_2-\omega_1)>0.
\end{equation}
where $C_i \equiv \kappa \rho_{i0} R_0^{n(1+\omega_i)}$ ($\rho_{i0}$ and $R_0$ denote initial values for the densities of both components and the radius of the shell). Provided $\omega_1 >-(n-1)/2n$, then \textit{the shell always become unstable for R large enough}.

Looking at (\ref{conservation}) we can see that at large $R$ the $\omega_1$-fluid dominates $\rho(R)$, so the asymptotic behaviour is determined by it 
and it can be shown that there are solutions where $R$ can take arbitrarily large values provided $\omega_1\geq0$. We are proving in this way that there exist shells that, while being solutions of Einstein equations, are unstable and therefore, as we will argue later on, unphysical.

\subsubsection*{\textbf{An example of a splitting shell}}
\label{splittingshells}

In this subsection we give an explicit solution of Einstein equations
where the splitting takes place at a critical point of stability condition (\ref{instability1}) (when the equality holds). This shell {\it becomes} unstable at that time, and a {\it soft} splitting solution may be built, that is, a splitting solution where $\ddot{R}$ is also continuous (which is what defines the critical condition).


We consider a shell made of two non-interacting fluids, whose parameters are as follows.

\begin{itemize}
\item $n=2$, $\kappa=8\pi$, $\omega_1=0$, $\omega_2=0.001$, $C_1=0.01$, $C_2=0.02$, $M_I=1$, $M_{II}=2$
\end{itemize}

We illustrate in Fig.2 the velocity of the shell and the stability condition as functions of $R$. There are no forbidden regions for the motion of this shell, so if it is initially expanding then it will continue to expand forever (provided the system evolves as a single shell during its entire evolution). We can see that at $R_c\approx 35,27$ the shell {\it becomes} unstable. If we consider an initial radius $R_0<R_c$, and say that the shell is initially expanding, then at $R=R_c$ we may construct a splitting of the type illustrated in Fig.1 where the $M_{int}$ parameter 
would be fixed by the continuity of $\dot{R}$.
As region $II$ must be exterior, and the energy densities of each component must be positive, the $r$ coordinate turns out to be a global coordinate (type $C$ splitting).

\begin{figure}
\label{ejemplofluidos}
\centerline{\includegraphics[height=7cm,angle=0]{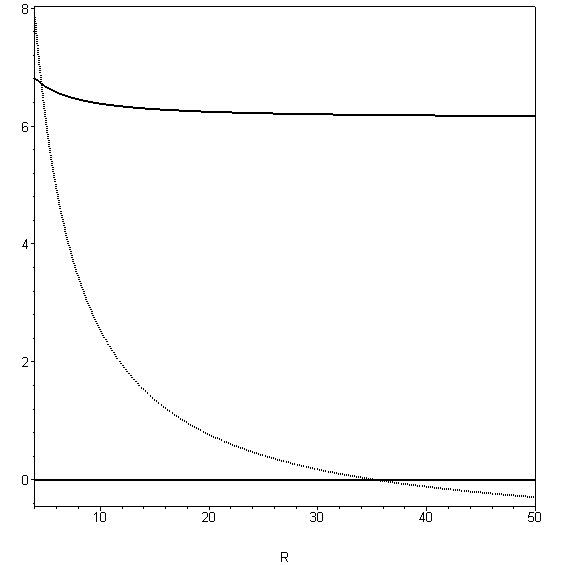}}
\caption{The solid curve represents $\dot{R}^2$ as a function of $R$ for a shell made of two non-interacting fluids with parameters $n=2$, $\kappa=8\pi$, $\omega_1=0$, $\omega_2=0.001$, $C_1=0.01$, $C_2=0.02$, $M_I=1$ and $M_{II}=2$. The dotted curve represents $1000\beta$. The critical point is located at $R\approx 35,27$; for $R$ greater than that the shell results unstable.}
\end{figure}

\subsubsection*{\textbf{Time-reversed splitting or merger of shells}}

It is noteworthy that the time-reversed picture of a splitting represents a collision of two shells whose outcome is a single one: a {\it merger} of shells. From an initial conditions perspective, the collisions that result from reversed splittings are far from arbitrary as they must satisfy that the relative velocity of the shells at the collision event is zero, that is, that their normal vectors coincide there. Particularly, in the context of shell collisions we affirm that our splitting analysis resembles one of the shell-crossing criteria used in the literature (see for instance \cite{NakaoIdaSugiura} and \cite{NuñezOliveira}), the so-called {\it transparency condition}, as both are motivated by the absence of any interaction other than gravity between the colliding shells. However, this {\it transparency condition} cannot resolve the outcome of the collision when applied to our colliding shells precisely because the collisions take place with zero relative velocity\footnote{In the notation of \cite{NakaoIdaSugiura}, we have $u_1^a n_{2a}=0$ at the collision. For these constructions, $u_1^a n_{2a}>0$ is imposed.}. Our infinitesimal separation analysis can be then considered an extension of the transparency condition for this limiting case. Although in this subsection we refer to time-reversed splitting, this should remain true for any pair of colliding shells intersecting with zero relative velocity: if the infinitesimal separation is stable then the outcome is a merger (a reversed splitting), but if it is unstable then the shells separate again and there would be a recoil or a crossing instead. Because of the continuity of $\dot{R}$, the criterion to determine the outcome must be related to the radial relative accelerations that the shells would have {\it right after} the collision, that is, when infinitesimally separated. Hence, the criterion would be precisely (\ref{instability1}): if $\beta>0$ the outcome is a merger, but if $\beta<0$ then there would be a shell crossing or a recoil. A more detailed analysis of this particular situation is outside of the scope of the present paper.

\section{Splitting brane-worlds}
\label{splittingbranes}

In this Section we extend the instability analysis that we made in the previous Section to actual brane-world scenarios.
We consider SMS brane-worlds \cite{SMS} in which the five-dimensional bulk spacetime has a spacelike 4-slicing in which every slice possesses a kind of 3-dimensional isotropy: spherical, planar or hyperbolic symmetry.

\subsection{Instability against separation of components in non-$Z_2$-symmetric scenarios}

We now propose a brane-world made of two non-interacting homogeneous and isotropic matter-energy components, so we may write $\rho=\rho_1+\rho_2$ and $p=p_1+p_2$. 
In this subsection we consider a splitting like that of Fig.1, and impose continuity of the normal vectors at the separation point. We assume that the whole spacetime possesses the symmetries we imposed above, so the intermediate region has to be Schwarzschild-AdS, with some $(M,\Lambda_5)$ parameters. As we assumed a unique cosmological constant for the entire spacetime, the geometry of the intermediate region would be determined by the mass parameter $M_{int}$ and the orientation of $\nabla r$ on it. 

In this scenario, equation (\ref{accelbrane2a}) can be applied to each splitting component in order to determine the relative acceleration of the separating shells. Taking the same steps as in Section \ref{2splitting}, we can see that the stability condition can also be calculated by (\ref{unstable}) and turns out to be precisely (\ref{instability1}) for $n=3$, which in this context can be written,
\begin{equation}
\label{instability2}
54(M_{II}-M_I)(\alpha_2-\alpha_1)-\kappa^2\rho^2R^4(3(\alpha_1+\alpha_2)+2) <0
\end{equation}
regardless of the relative orientation of $\nabla r$ in the intermediate region. Now, if we consider that the brane tension stress-energy is indivisible, one of the two splitting shells should contain brane tension. In this way, we may write $\alpha_1=(-b+p_{1m})/(b+\rho_{1m})$\footnote{Because both regions $I$ and $II$ are {\it interior}, a permutation of the labels $1$ and $2$ is equivalent to a permutation of $I$ and $II$.}. In the low-energy limit $\alpha_1$ simply becomes $\alpha_1\approx -1$, while the stability condition would take the form  
\begin{equation}
\label{2braneins}
\kappa^2b^2R^4(3\alpha_2-1)-54(M_{II}-M_I)(\alpha_2+1) > 0.
\end{equation}
Because there are very tight observational bounds on dark radiation, this stability condition for ``realistic'' brane-world scenarios turns out to be merely $\alpha_2>1/3$, which is a condition that matter made of particles does not satisfy. We may consider ($\rho_2,p_2$) as the matter-energy content of the universe, or some non-interacting part of it (like the dark matter component to a good approximation), which is thought to have always satisfied $\alpha<1/3$, and, indeed, it satisfies $\alpha\approx 0$ in the matter-dominated era.
We then conclude that {\it non-$Z_2$-symmetric brane-world scenarios are typically unstable against separation of non-interacting components}.

Although the name of this subsection makes reference to non-$Z_2$-symmetric scenarios, this kind of instability would also hold for an {\it initially} $Z_2$ symmetric bulk spacetime. Condition (\ref{2braneins}) makes perfect sense if $M_I=M_{II}$, which is equivalent in this context to having $Z_2$ symmetry. In that case, a two-component splitting would necessarily break the symmetry (otherwise the brane tension component must remain at the center of symmetry), so if we have a reason to impose it, as it may happen in brane-world models, such a splitting scenario should not be taken into account.

\subsection{Instability against separation of a three-component brane-world in a $Z_2$-symmetric scenario}

In the context of brane cosmology, a common simplifying assumption coming from string theory is the existence of $Z_2$ symmetry for the bulk spacetime, such that the brane-world is the set of fixed points of this symmetry \cite{RS2}.
As we pointed out, an analysis involving a fragmentation into two parts can not be possible if this symmetry holds. Nevertheless, nothing prevents us to consider a different kind of splitting for the $Z_2$ symmetric scenario, as illustrated in Fig.3. To preserve the symmetry, we propose a fragmentation into three parts, one of which remains at the symmetry center, while the other two, being mirror images of one another, move away from the set of fixed points.

\begin{figure}
\label{splitting3}
\centerline{\includegraphics[height=8cm,angle=0]{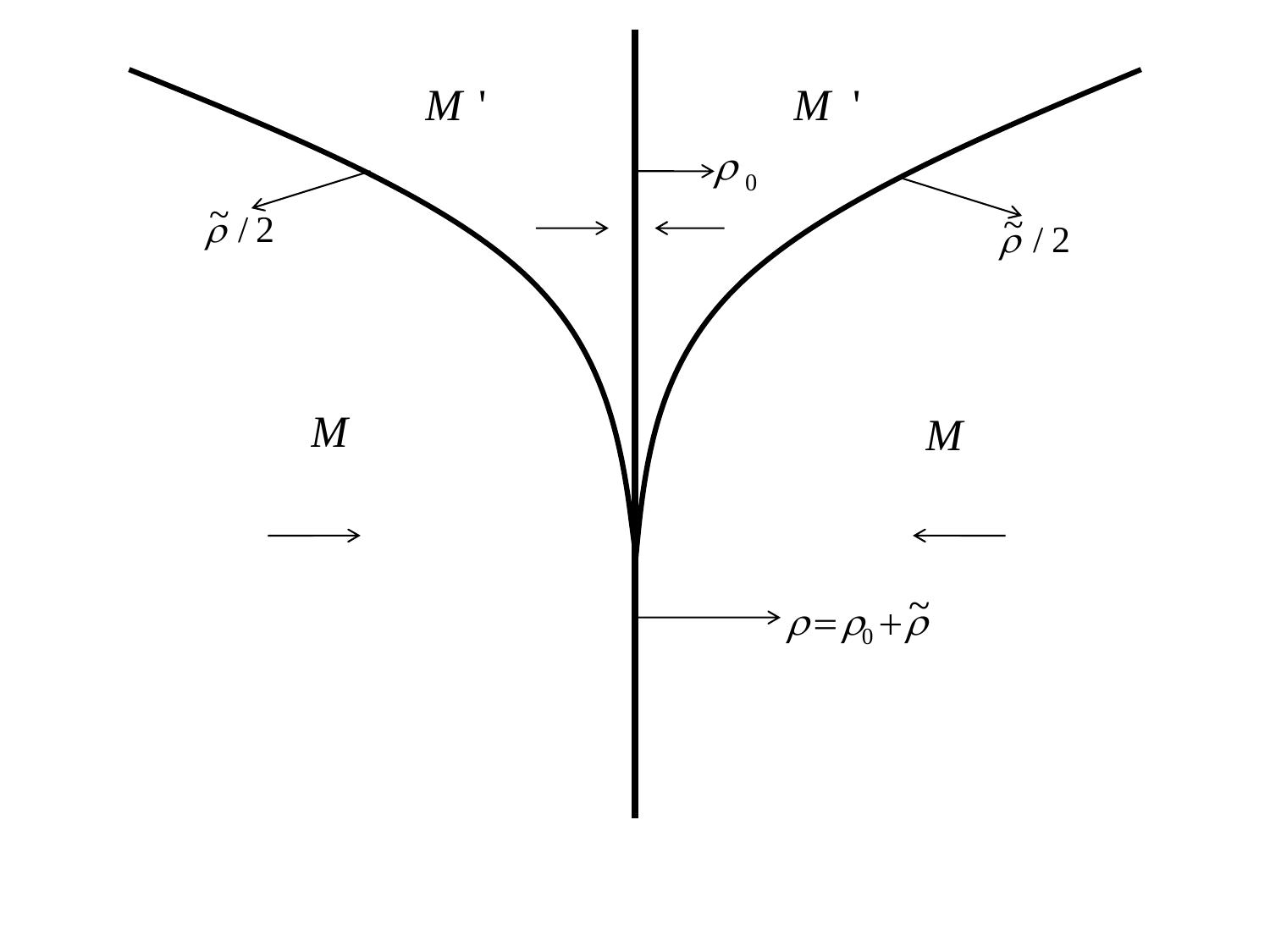}}
\caption{Schematic representation of a {\it splitting} into three parts that preserves $Z_2$ symmetry. The arrows in each empty region represent the orientations of $\nabla_a r$.}
\end{figure}

This is a three-component (non-interacting) shell splitting, where two of them are identical. We can consider the ``outgoing'' shells as two non-interacting identical parts of a single matter-energy component, whenever such interpretation makes sense (as in the case of dust).
Analogously to Section \ref{2splitting}, the key to analyze this kind of instability is to use (\ref{accel1}) in order to obtain the relative acceleration of the resulting shells. We call $M$ the mass parameter of the Schwarzschild-AdS regions before the splitting, and $M'$ the mass parameter that would correspond to any of the regions between the symmetry center and one of the ``outgoing'' shells. We now define $\rho_0$ and $p_0$ as the stress-energy tensor components of the central splitting shell, and denote the corresponding components for the outgoing shells by $\tilde{\rho}/2$ and $\tilde{p}/2$, so we can write $\rho=\rho_0+\tilde{\rho}$ and $p=p_0+\tilde{p}$. The equation of motion of the resulting central shell is obtained by replacing $(M_I,M_{II},\rho(R))$ in (\ref{BranesinZ2}) with $(M',M',\rho_0(R))$; whereas for the outgoing shells it is obtained by replacing those quantities in the same equation with $(M',M,\tilde{\rho}(R)/2)$.
Continuity of $\dot{R}(\tau_0)$ determines $M'$ as a function of $(M,R_0)$ and the matter models parameters, and it can be shown that $M'$ is always well-defined. Positivity of each energy density component implies that $\nabla_a r$ always points to the central brane, which in turn implies $M'>M$.

We have for each shell at the splitting time,
\begin{eqnarray}
\ddot{\tilde{R}}(\tau_0) &=&-\frac{1}{R_0}\left[k+\frac{2R_0^2}{\ell^2}+\dot{R}^2(\tau_0)-\left(3\tilde{\alpha}+1\right)\sqrt{\dot{R}^2(\tau_0)+F}
\sqrt{\dot{R}^2(\tau_0)+F'} \right] \nonumber \\
\label{accelsplit2}
\ddot{R_0}(\tau_0) &=& - \frac{3\alpha_0+2}{R_0}\left(\dot{R}^2(\tau_0)+ F'\right) -\frac{2M'}{R_0^3}-\frac{R_0}{\ell^2}
\end{eqnarray}
and, analogously to Section \ref{2splitting}, the condition for having stability would be simply
\begin{equation}
\ddot{\tilde{R}}(\tau_0) - \ddot{R_0}(\tau_0)>0.
\end{equation}
After manipulating (\ref{accelsplit2}) and taking into account that $\rho>0$ and $\rho_0>0$, we finally get
\begin{equation}
\label{3braneins}
(3\alpha_0+1)\rho_0+(3\tilde{\alpha}+1)(\tilde{\rho}+\rho_0)>0.
\end{equation}
This inequality can also be written as
\begin{equation}
\label{instability3}
3(p_0+\rho_0)+(3\tilde{\alpha}-1)\rho_0+(3\tilde{\alpha}+1)\tilde{\rho}>0,
\end{equation}
which makes explicit that {\it if $\tilde{\alpha} \geq 1/3$, then the shell is always stable}.
We now suppose that on the original shell there is brane tension plus matter satisfying the dominant energy condition, so we may write $\rho_0=b+\rho_{0m}$, $p_0=-b+p_{0m}$, which illustrates the fact that the resulting central shell must contain the brane tension\footnote{If we consider that the brane tension component is indivisible, then it must remain at the symmetry center in order to satisfy the junction conditions.}.
In this case, we can rewrite the inequality (\ref{instability2}) as follows,
\begin{equation}
(3(\alpha_{0m}+\tilde{\alpha})+2)\rho_{0m}+(3\tilde{\alpha}-1)b+
(3\tilde{\alpha}+1)\tilde{\rho}>0
\end{equation}
where it can be seen that in the low-energy limit the stability condition is simply $\tilde{\alpha} \geq 1/3$, which is almost the same condition than that for a two-component splitting. The difference lays in the fact that if the separating component is an ensemble of non-interacting massless particles then the configuration is stable, unlike the case of a two-component splitting in a non-$Z_2$-symmetric setting. On the other hand, matter made of massive particles always satisfies $\tilde{\alpha} < 1/3$, so if it makes sense to separate that component into two identical non-interacting parts, as in the cases of dust and Vlasov matter, then the system can be considered unstable.


In this way, the dark matter component of the universe, if thought as an ensemble of non-interacting particles, can always be the separating component of this scenario, so in this sense we can say that {\it $Z_2$-symmetric brane-world scenarios are typically unstable against separation of the dark matter component into the bulk}.

\subsection{A ``realistic'' example}

In this subsection we want to illustrate this instability of typical $Z_2$-symmetric brane-worlds in the context of actual brane-world models considered in the literature. This will also serve as an example of how to construct a splitting solution in this context, which might illustrate some of the possible ``phenomenological'' (potentially observable) consequences of living in a brane-world that splits into several parts. We consider a SMS brane-world universe whose {\it effective} matter-energy content resembles the $\Lambda$-CDM universe.
We may write,


\begin{equation}
\rho_m=\rho_c (\Omega_c+\Omega_b)(1+z)^3+\epsilon^{\gamma}_0(1+z)^4
\end{equation}
where $\rho_c\equiv 3H_0^2/8 \pi G$ is the critical density at present cosmological time, $\Omega_c$ and $\Omega_b$ are the relative proportions of dark matter and baryonic matter respectively at present with respect to the critical density, and $\epsilon^{\gamma}_0$ is the present radiation energy density. We consider the simplest choice of parameters ($k=M=0$) for both empty bulk regions, which particularly implies non-existence of ``dark radiation'' and $Z_2$ symmetry (each bulk region is an identical piece of $AdS_5$). The Friedmann equation (\ref{FriedmannsinZ2}) for such a universe reads,
\begin{equation}
\label{Friedmann0a}
H^2=H_0^2\left[\left((\Omega_c+\Omega_b)(1+z)^3+\frac{\epsilon_0^{\gamma}}{\rho_c}(1+z)^4\right)\left(1+\frac{\epsilon_0^{\gamma}(1+z)^4}{2b}\right)+\Omega_{\Lambda}\right]
\end{equation}
where $\Omega_{\Lambda}\equiv \Lambda_4/(3H_0^2)$, and we have ignored every term other than the radiation term in the quadratic in $\rho_m$ part of this expression. For small $z$, this equation takes the standard form,
\begin{equation}
\label{Friedmann0}
H^2=H_0^2((\Omega_c+\Omega_b)(1+z)^3+\Omega_{\Lambda})
\end{equation}
which makes explicit the relation $\Omega_c+\Omega_b+\Omega_{\Lambda}=1$.

{\it We propose a splitting into three parts like that of Fig.3, where the $(\tilde{\rho},\tilde{p})$ component is the dark matter component $(\tilde{\rho}=\rho_c\Omega_c(1+z)^3,\tilde{p}=0)$ and the splitting takes place at present cosmological time ($z=0$)}. It makes sense to separate two identical dark matter components, which would constitute the two ``outgoing'' shells, as cold dark matter is supposed to be made of non-interacting particles, hence it can be modeled as Vlasov matter, and at cosmological scales simply as dust. We should emphasize however that we do not expect this kind of construction to have {\it a priori} any degree of correspondence with the actual universe.
They are solutions of Einstein equations that correspond to initial value data that can be obtained from a family of brane-world models analyzed in the literature. As actual solutions, we may consider the phenomenology that one would expect for observers living in a splitting brane-world, just in order to explore the theoretical possibility. In any case, with this construction we provide another example of the {\it lack of uniqueness} in the evolution of initial data that corresponds to concentrated sources.


Replacing $\rho$ with $\tilde{\rho}/2$ and setting $(M_I=0,M_{II}=M')$ in (\ref{BranesinZ2}), the equation of motion for any of the ``out-going'' dust shells can be written as follows,
\begin{equation}
\tilde{H}^2=\frac{8 \pi G}{3}\tilde{\epsilon}_{dr}\left(1+\frac{2b\tilde{\epsilon}_{dr}}{\tilde{\rho}^2}\right)+\frac{\pi G}{3b}\tilde{\rho}^2+\frac{\Lambda_4-4\pi Gb}{3}
\end{equation}
where we replaced $\kappa$ and $\Lambda_5$ as functions of $G$ and $b$, and $8\pi G\tilde{\epsilon}_{dr}/3\equiv M'/R^4$ is the ``dark radiation'' term associated with the mass parameter $M'$ of the intermediate region. Taking into account the fact that $\tilde{\rho}<<b$ at the separation moment, we can write

\begin{equation}
\label{Friedmann1a}
\tilde{H}^2=H_0^2\left[\tilde{\Omega}_{dr}(1+\tilde{z})^4 \left(1+\frac{2b \tilde{\Omega}_{dr}}{\rho_c \Omega_c^2(1+\tilde{z})^2} \right)+\Omega_{\Lambda}-\frac{b}{2\rho_c}\right]
\end{equation}
where $\tilde{\Omega}_{dr}\equiv \tilde{\epsilon}_{dr} (R_0)/ \rho_c$ and $\tilde{z}\equiv R_0/\tilde{R}-1$. In this way, all we need to determine the equation of motion of the resulting dust shells is the parameter $\tilde{\Omega}_{dr}$, which can be found by continuity of the normal vectors at the separation point, that is, by solving it from $H(z=0)=\tilde{H}(\tilde{z}=0)$. From (\ref{Friedmann0}) and (\ref{Friedmann1a}) we get,
\begin{equation}
\tilde{\Omega}_{dr}^2 \approx \frac{\Omega_c^2}{4}\left(\frac{\rho_c}{b}(2\Omega_b+\Omega_c)+1\right)
\end{equation}
so (\ref{Friedmann1a}) can be written
\begin{equation}
\tilde{H}^2=H_0^2\left[\left(\frac{b}{2\rho_c}+\frac{\Omega_c}{2}+\Omega_b\right)(1+\tilde{z})^2 +\frac{\Omega_c}{2}(1+\tilde{z})^4+\Omega_{\Lambda}-\frac{b}{2\rho_c}\right].
\end{equation}
This expression can be simplified neglecting terms ($b>>\rho_c$), then we finally get,
\begin{equation}
\label{Friedmann1b}
\tilde{H}^2=H_0^2\left[1+\frac{b}{\rho_c}\tilde{z}\left(1+\frac{\tilde{z}}{2}\right)\right]
\end{equation}
where it can be easily seen that the dust shells almost immediately rebound ($\tilde{H}=0$ at $\tilde{z}\approx -\rho_c/b$), which means that the scale parameter $\tilde{R}$ begins to decrease until they collapse and form a pair of bulk black holes with mass parameter $M'$.

On the other hand, the equation of motion of the central shell after the splitting can be computed from (\ref{FriedmannsinZ2}) by writing $\rho_m=\Omega_b(1+z)^3$ and $M_{II}=M_I=M'$,
\begin{equation}
\label{Friedmann2}
H^2=H_0^2\left[\Omega_b(1+z)^3+\Omega_c(1+z)^4 +\Omega_{\Lambda} \right]
\end{equation}
where the second term of the right hand side corresponds to the ``dark radiation'' term $M'/R^4$. In this way, the motion of the central shell is determined by (\ref{Friedmann0}) for $z>0$ and by (\ref{Friedmann2}) for $z<0$. As we can see, the time derivative of the Hubble parameter would be discontinuous for the central shell, and the dynamical effect of the ``disappearance'' of the dark matter component would be {\it as if the dark matter of the universe transformed into dark radiation}, as the initial ($z=0$) effective energy density of dark radiation is precisely the dark matter density ($\rho_c\Omega_c$) at the time of the splitting.

We end this Section with the remark that the critical point for any of the stability conditions (\ref{2braneins},\ref{3braneins}) takes place at very early stages of the universe, long before nucleosynthesis.
In this way, the brane results unstable during almost its entire evolution, and becomes increasingly unstable as time passes. A ``soft'' splitting, like those illustrated in Section \ref{splittingshells}, where $\ddot{R}$ is continuous, can only take place at the critical point of the stability condition, so it could only happen in the inflationary era (taking into account experimental bounds on the brane tension based on deviations of Newtonian gravitation at submillimeter scales \cite{LCCGVP}), or even before. We may speculate in further investigations the phenomenological consequences of having this kind of splitting in the context of the very early universe.

\section{Discussion}
\label{final}


\subsection{Splitting, brane collisions and conservation laws}


As far as we know, this particular kind of phenomena in the context of brane-world cosmologies was not specifically considered in the literature. Nevertheless, more general scenarios have been studied in relation to brane-world collisions. Langlois, Maeda and Wands considered in \cite{LangloisMaeda} an arbitrary number of intersecting hypersurfaces where the common intersection is a codimension $2$ manifold, and obtained a {\it conservation law} that is a requirement for the absence of a conical singularity at the intersection. Moreover, Gravanis and Willison \cite{GravanisWillison} considered general networks of singular hypersurfaces in Lovelock gravity, developed a formalism to define connections on this spacetime, and obtained selection rules for collisions provided the matter-energy satisfies an energy condition. Our constructions can be thought of as some of the simplest cases among these networks of thin shells, with the further restriction of {\it differentiability} at the so-called {\it collision event}.
Anyway, the physical motivation in this work is completely different (or much more specific) from the analysis of brane collisions, and it is precisely this motivation what requires the ``extra'' differentiability assumption: we interpret the ``collision event'' as a {\it perturbation}, so it should be as innocuous as possible, and it can be merely thought of as a spontaneous infinitesimal separation of non-interacting constituents.

In particular, as a special case of the constructions made in \cite{LangloisMaeda}, the splitting shells we considered comply with stress-energy conservation in the same sense of that paper. Our constructions satisfy the ``conservation law'' (Eq. $16$ of that paper) that Langlois {\it et.al.} found, which in our case turns out to be equivalent to the ``continuity'' of the stress-energy tensor\footnote{Every Lorentz factor in Eq. $18$ of \cite{LangloisMaeda} is either $1$ or $-1$ as a consequence of the continuity of $\dot{R}$.}. With ``continuity'' we mean that   
the stress-energy tensor at the bifurcation submanifold is precisely the sum of the stress-energy tensors of the separating shells. This last statement would not make sense in a general collision because the stress-energy tensors of the colliding shells are defined on different tangent spaces. In our case, and as a consequence of the continuity of the normal vector, the tangent spaces of each submanifold coincides at the intersection, so there we can unambiguously operate tensors defined at any of the separating submanifolds.


On the other hand, if we understand splitting processes as matter-energy leaving a brane (we could define {\it the brane} as the shell in which the brane tension lives), we should take into account that there might be reasons coming from string theory to {\it a priori} preclude any chance of having matter moving off into the bulk, as one may consider branes as fundamental objects by themselves, which appear in the definition of the action that describes the theory one is seeking to build. We may consider branes as objects {\it a priori} defined in some sense. Indeed, the networks of intersecting hypersurfaces in \cite{GravanisWillison} are interpreted in this sense, so there is no discussion regarding the {\it evolution} of the topology of the network: it is already {\it given} and the theory itself accommodates to it.
We then basically affirm that the brane-world models considered in our work, in the framework of General Relativity defined on the entire {\it bulk} spacetime without any {\it a priori} ``special'' object, are unstable, or at least they entail non-uniqueness in the evolution (we refer to this later in this Section), and the existence of those features may be an argument against the plausibility of these systems, and of certain families of thin shell models in general.

\subsection{Constraints on the ``microphysics''}

 We could also interpret the instability results as evidence for the existence of some mechanism that would ``prevent'' the splitting, that is, some interaction between the constituents and a external field, or among the constituents themselves, that may compensate the gravity-driven instability. For example, we may take into account a scalar field on the bulk, which is a widely considered scenario as it may naturally exhibit an inflationary stage and it can be justified by string theoretical considerations. Another possibility is to consider Lovelock gravity in the bulk. These are possible future research directions. It may well happen that some brane-world scenarios turn out to be stable in this sense, and the stability against separation of constituents could be an argument in favour of some of them. 


\subsection{The initial value problem with concentrated sources}

As mentioned, another interesting issue related to the splitting solutions is the {\it lack of uniqueness} in the evolution of surface layers. We have seen different situations involving singular hypersurfaces where a splitting solution can be constructed. But the splitting process is something that we can not acknowledge by looking at initial data on a Cauchy surface that intersects the thin shell before the separation point. For every splitting solution that we constructed in this work, there is also a corresponding non-splitting solution where the shell moves according to an equation of motion, and it may collapse, oscillate or expand without any fragmentation whatsoever. The corresponding initial data in one of these Cauchy surfaces is exactly the same for the splitting solution as for the non-splitting solution, which means that the initial value problem for these objects is compromised. Moreover, there are generally an infinite number of splitting solutions corresponding to a given unstable thin shell, as there is freedom in choosing the separation moment (the time lapse of unstable evolution) and sometimes also in choosing the pair of separating non-interacting components.
Further, if we consider particle evaporation we could also build solutions where an {\it atmosphere} forms, creating even more possible evolutions. 

This may not be surprising taking into account that we are dealing with {\it distributional} initial data. In other contexts involving hyperbolic PDE, like ideal fluids in Minkowski spacetime, surfaces of discontinuity (shock-waves) may form and evolve, and their evolution according to the original PDE system may not be unique \cite{Christodoulou}. Anyway, there is an external principle (an entropy law) that recovers uniqueness. In this context we may have an analogous situation: one may identify the ``physical'' evolution among the mathematical possibilities by means of some principle arising from the ``microphysics'' of the matter-energy content of the thin shell. After all, the information about the splitting might be contained exclusively in the degrees of freedom associated with the constituent matter fields of the singular surface. A specific treatment of these issues should be considered in a separate paper.

\section{Concluding Remarks}
\label{concluding}

Summarizing, we have shown that the stability against separation analysis developed in \cite{paper1} naturally generalizes for general non-interacting matter fields. In section \ref{2splitting} we obtained a general criterion ($\ref{instability1}$) for spherically symmetric shells in $D$ dimensions which is an inequality involving the energy densities and pressures of the non-interacting constituents as well as the shell parameters. We found that the instability essentially takes place when the difference between the $\alpha$ parameters of the components (the pressure to energy density ratio) is large enough. Later, in section \ref{splittingbranes}, we extended this instability analysis to $5$-dimensional Schwarzschild-AdS spacetimes in the context of SMS brane-world cosmology. For a splitting into two shells we found the same criterion ($\ref{instability2}$) as for the spherically symmetric case with $n=3$. When applied to a brane-world context, this criterion essentially implies that any realistic matter field in a simple SMS brane-world model would separate from the brane tension component. Then we performed a different stability analysis adapted to a $Z_2$-symmetric universe: a splitting into three parts. We obtained the criterion ($\ref{instability3}$), which also implies that any realistic matter field component, with the exception of a gas of massless particles, would separate from the brane tension component. Finally, we developed an example of a splitting brane-world adapted to observed cosmological parameters, and illustrate some aspects of the phenomenology that might arise for this kind of solutions. 


As a final comment, looking at the results in \cite{paper2} where it is shown that certain singular surfaces are indeed thin shell limits of thick configurations, and that all those singular surfaces are stable against separation of constituents, we make the following conjecture: {\it unstable shells are not thin shell limits of thick configurations.} If we consider thin shells as idealizations of ultimately thick configurations, those unstable solutions could be interpreted as misrepresentations of thick configurations, and hence they can only be taken into account if we consider the system to be {\it fundamentally thin}, that is, if there is some fundamental principle that simply precludes the possibility of having matter-energy moving off the surface. An unstable thin shell may be considered as a {\it spurious solution} arising from the loss of information associated with the fact that we neglected a spatial dimension. A general demonstration of this conjecture constitutes a possible future research direction, and it should be deeply related to the general initial value problem for concentrated sources. In any case, the stability-against-separation analysis that we exposed in this work can be regarded as a novel analysis to be taken into account in order to constrain the plausibility of models involving surface layers.

\section{Acknowledgments}

This work was partly supported by CONICET. I wish to thank Reinaldo Gleiser for inspiration, insightful ideas, and for reading the whole manuscript.  I also thank Hideki Maeda for reading the manuscript, Robert Geroch for illuminating discussions and advice, and Oscar Reula for useful discussions regarding the initial value problem for shock waves. MR is supported by CONICET.

\end{document}